\documentclass[12pt]{article}
\usepackage{a4,amssymb,verbatim,amsmath}
 \usepackage{color}

\newtheorem{theorem}{Theorem}[section]

\newtheorem{remark}[theorem]{Remark}

\newtheorem{ex}{Example}[section]
\newenvironment{example}{\begin{ex}\rm}{ \hfill $\Diamond$ \end{ex}
        \vskip4pt}
\newtheorem{ass}{Assumption}[section]

\numberwithin{equation}{section}

\begin{document}

\def\sso#1{\ensuremath{\mathfrak{#1}}}  
\def\ssobig#1{{{#1}}}

\newcommand{\ddx}{\partial \over \partial x}
\newcommand{\ddy}{\partial \over \partial y}

\begin{center}
{\bf \Large 
Linear or linearizable first-order delay ordinary differential equations 
and their Lie point symmetries }
 \end{center}

\bigskip

\begin{center}
{\large Vladimir A. Dorodnitsyn}$^{*}$,
{\large Roman Kozlov}$^{**}$, \\
{\large  Sergey V. Meleshko}$^{\dag}$
{\large and Pavel Winternitz}$^{\ddag}$

\bigskip

${}^{*}$
Keldysh Institute of Applied Mathematics, Russian Academy of Science, \\
Miusskaya Pl.~4, Moscow, 125047, Russia; \\
e-mail address: DorodnitsynVA@gmail.com \\

$^{**}$ Department of Business and Management Science, \\ Norwegian
School of Economics, Helleveien 30, 5045, Bergen, Norway; \\
 {e-mail: Roman.Kozlov@nhh.no}\\

${}^{\dag}$
School of Mathematics, Institute of Science, \\ Suranaree University of Technology, 30000, Thailand; \\
e-mail address: sergey@math.sut.ac.th \\

$^{\ddag}$ Centre de Recherches Math\'ematiques
and
D\'epartement de math\'ematiques et de statistique,
Universit\'e de Montr\'eal,
Montr\'eal, QC, H3C 3J7, Canada; \\
e-mail address: wintern@crm.umontreal.ca \\

\end{center}

\bigskip

\begin{center}
\end{center}

\bigskip

\begin{center}
{\bf Abstract}
\end{center}
\begin{quotation}
A previous article was devoted to an analysis of the symmetry properties 
of a class of first-order delay ordinary differential systems (DODSs). 
Here we concentrate on linear DODSs. 
They have infinite-dimensional Lie point symmetry groups due to the linear superposition principle. 
Their symmetry algebra always contains a two-dimensional {sub}algebra realized by linearly connected vector fields. 
We identify all classes of linear first-order DODSs that have  additional symmetries, 
not due to linearity alone. 
We present representatives of each class. 
These additional symmetries are then used to construct exact analytical particular solutions using symmetry reduction.
\end{quotation}

\eject 

\section{\large \bf  Introduction}     \label{ Introduction}

The recent article~\cite{ourpaper} was devoted to the Lie group classification
 of first order delay ordinary differential system (DODS). 
For motivation and brief survey of the field we refer to~\cite{ourpaper} and references therein.

The considered DODS consists of a pair of equations
\begin{equation}  \label{DODE}
\dot{y} = f ( x, y, y_- ), 
\qquad 
x_- = g ( x, y, y_- )   , 
\end{equation}
$$
x \in I , 
\qquad 
{ \partial f  \over  \partial y_-   }   ( x, y, y_- )  {\not\equiv}  0  , 
\qquad 
x_- <  x , 
\qquad g  ( x, y, y_- ) {\not  \equiv}   \mbox{const} ,
$$
where $I \subset  \mathbb{R} $ is some finite or semi{finite}
interval. Functions $f$ and $g$ are locally smooth functions in some domain 
$\Omega   \subset  \mathbb{R}^3 $. 
For the first equation in equation~(\ref{DODE})  we have to specify the delay point $x_-$
where the delayed function value $ y_- = y(x_-) $ is taken,
otherwise the problem is not fully determined.

In~\cite{ourpaper} DODS were classified into equivalence classes under
 the local group of diffeomorphisms (the equivalence group)
\begin{equation}    \label{equivalence}
\bar{x} = \bar{x} ( x, y) ,
\qquad
\bar{y} = \bar{y} ( x, y) .
\end{equation}
Each class is represented by a DODS with a specific Lie point symmetry algebra. The symmetry algebras are represented
by vector fields.

We are interested in DODSs which are invariant with respect to
point symmetry groups.
Such groups consist of transformations generated by the vector fields
of the same form as in the case of ordinary differential equations
\cite{Ovsiannikov1982, Olver1986, Ibragimov1985, Bluman1989, Gaeta1994},
namely
\begin{equation}    \label{operator1}
X _{\alpha}  =   \xi  _{\alpha}   (x,y)  { \ddx}
+ \eta  _{\alpha}  (x,y)  { \ddy}  ,
\qquad
\alpha  = 1, ..., n .
\end{equation}
The prolongation of these vector fields acting on the system~(\ref{DODE})
has the form
\begin{equation}    \label{prolongation}
\mbox{\bf pr} X  _{\alpha}
=    \xi _{\alpha}  { \ddx}  + \eta _{\alpha}   { \ddy}
+   \xi  _{\alpha}   ^-   {\partial  \over \partial x_-}  + \eta _{\alpha}   ^-  {\partial   \over \partial y_-}
+   \zeta  _{\alpha}   {\partial  \over \partial \dot{y} }
\end{equation}
with
\begin{equation*}
\xi _{\alpha}   = \xi  _{\alpha} (x,y) ,
\qquad
\eta _{\alpha}  = \eta   _{\alpha}  (x ,y )  ,
\end{equation*}
\begin{equation*}
\xi _{\alpha} ^-    = \xi   _{\alpha}  (x_- ,y_-)  ,
\qquad
\eta  _{\alpha} ^-    = \eta   _{\alpha} (x_- ,y_-)  ,
\end{equation*}
\begin{equation*}
\zeta   _{\alpha}   (x,y, \dot{y} )  = D  ( \eta  _{\alpha}  )  - \dot{y}   D  ( \xi   _{\alpha} )  ,
\end{equation*}
where $D$ is the total derivative operator.
The operators~(\ref{prolongation}) combine
prolongation for shifted discrete variables $(x_- , y_-)$~\cite{DKW2000, Dorodnitsyn2011, Levi2006}
with standard prolongation for  the  derivative $\dot{y}$~\cite{Ovsiannikov1982, Olver1986}.

In~\cite{ourpaper} the main emphasize was on {\it nonlinear} DODS, on their Lie point symmetry algebras,
and on exact solutions obtained by symmetry reduction. 
In particular it was shown that genuinely nonlinear DODS can have symmetry algebras of dimension
$\mbox{dim} \  L =n$ with $n=0, 1, 2, 3$. 
By "genuinely nonlinear" we mean nonlinear DODS that cannot be linearized by a point transformation.

Two significant results concerning linear DODS were also obtained
in~\cite{ourpaper}, namely the following. We consider the most
general DODS of the form~(\ref{DODE})  with solution independent
delay point $x_- $:
\begin{equation}    \label{linearDODE}
 \dot{y} = \alpha  ( x  )  y +   \beta  ( x  )  y_-  +  \gamma  ( x  )  ,
\qquad
x_-   = g  ( x )   ,
\end{equation}
where $\alpha (x)$, $ \beta (x) $,  $ \gamma (x) $ and $ g(x) $
are arbitrary real functions,
smooth in some interval $ x \in I $.

From now on we  use the term {\it linear } DODS for systems of the form~(\ref{linearDODE}). 
The symmetry algebra of~(\ref{linearDODE}) is infinite-dimensional
for all functions $\alpha (x)$, $ \beta (x) $,  $ \gamma (x), $ and $
g(x) $. The reason for this results is quite simple. In order to
solve the DODS~(\ref{DODE}) (in particular~(\ref{linearDODE}))
we must give some initial conditions (see
\cite{Elgolts1964,bk:Myshkis1972}). Contrary to the case of ordinary
differential equations, the initial condition must be given by a
function $ \varphi (x)$ on an initial interval $ I_0 \subset
\mathbb{R} $, e.g.
\begin{equation}    \label{initialvalues}
 y (x)   = \varphi (x), \qquad
x \in [ x_{-1} , x_0 ]  .
\end{equation}
Here it is assumed that $ x_{-1} = g ( x_0 ) $.

For linear DODS the freedom in the choice of the initial function
$\varphi (x)$ is reflected in a linear superposition formula. This
is formulated as a theorem of~\cite{ourpaper}: 
The linear DODS~(\ref{linearDODE}) admits  an infinite-dimensional Lie algebra
represented by the vector field:
\begin{equation}    \label{always}
X (\rho)  =  \rho  (x)  { \ddy} , \qquad Y (\sigma)  =  ( y  -
\sigma  (x) )  { \ddy}  ,
\end{equation}
where $ \rho (x) $  is the general solution of the homogeneous
DODS obtained by putting  $\gamma (x)=0$ in~(\ref{linearDODE}) 
and  $ \sigma (x) $ is
any  chosen particular solution of the inhomogeneous
DODS~(\ref{linearDODE}). 
The fact that~(\ref{always}) is a symmetry algebra of  the DODS~(\ref{linearDODE})  
is a consequence of linearity alone. 
We shall call this algebra $ \mbox{\bf A}_{\infty} $. 
As we shall see below, for some special cases of~(\ref{linearDODE}) 
the symmetry algebra can be larger.

The infinite-dimensional Lie algebra of~(\ref{linearDODE}) always
contains a two-dimensional subalgebra realized by  linearly
connected vector fields. Two possibilities exist:

1.  The solvable nonnilponent algebra  a basis of which can be
transformed into $ \mbox{\bf A}_{2,1} $:

\begin{equation}      \label{Dcase24}
X_1 = { \ddy}   ,
\qquad
X_2 =  y  { \ddy}.
\end{equation}
The invariant DODS is homogeneous and given by 
\begin{equation} \label{DODEcase24}
\dot{y}   =    f (x)   {  \Delta y     \over    \Delta  x   } ,
\qquad .
   x_-    =      g (x)    ,
\qquad
   g (x) < x ,
\qquad
 g(x) {\not\equiv}  \mbox{const}    .
\end{equation}

2. An Abelian Lie algebra $ \mbox{\bf A}_{2,3} $ a basis of which can be
transformed into
\begin{equation}      \label{Dcase22}
X_1 = { \ddy}   ,
\qquad
X_2 = x  { \ddy}.
\end{equation}
The invariant DODE is inhomogeneous and the DODS can be presented as
\begin{equation}       \label{DODEcase22}
\dot{y}  =   {  \Delta y    \over   \Delta  x  }    +  f (x)   ,
\qquad
   x_-    =     g ( x )   ,
\qquad
   g (x) < x ,
\qquad
 g(x) {\not\equiv}  \mbox{const}    .
\end{equation}
We  assume that $ f(x) {\not\equiv} 0 $, as
otherwise~(\ref{DODEcase22}) is a special case of~(\ref{DODEcase24}).

\medskip

{\bf Comment.}  If we know at least one solution $\sigma_0(x)$ of
inhomogeneous equation ~(\ref{DODEcase22}), then we can transform it into a homogeneous
DODS by the transformation 
$\bar{x} =x$,  $ \bar{y} = y - \sigma_0 (x) $.  
However, it is not assumed 
that such a solution ${\sigma_0(x)}$ is known.

\medskip

{\bf Comment.}  To explain the notations we note that two more inequivalent
two-dimensional algebra of vector field exist, represented by
\begin{equation} \label{int13}
 \begin{array}{cll}
{ \displaystyle
  \bf A_{2,2} }: & { \displaystyle  X_{1} = { \ddy},}
 & { \displaystyle X_{2} = x{ \ddx}+ y{ \ddy} ;  } \\
 \\
 { \displaystyle
   \bf A_{2,4} }: &{ \displaystyle  X_{1} = { \ddy}, }
  & { \displaystyle X_{2} = { \ddx} . }  \\
 \end{array}
 \end{equation}
The vector fields $X_{1}$ and $X_{2}$ are linearly {non}connected, and
therefore the invariant DODS are nonlinear 
and can not be transformed into a linear one.

\medskip

The purpose of this article is to provide a classification of linear
DODS of the form~(\ref{linearDODE}), and to obtain solutions which 
are invariant under some one-dimensional {sub}algebra. 
To do this we analyze the finite-dimensional {sub}algebras  
of the infinite-dimensional symmetry algebras of linear DODS. 
It terns our that it is sufficient to consider dimensions 
$ 2 \leq n \leq 4 $ to fully characterize a class of equations. 
We already
know that the finite-dimensional  {sub}algebras always contain
the algebra ~(\ref{Dcase24}) or~(\ref{Dcase22}), possibly both. The presence of
further vector fields  simply imposes restrictions on the
functions $f(x)$ and $g(x)$ in ~(\ref{DODEcase24}), and~(\ref{DODEcase22}).

The article is organized as follows. In Section~\ref{generaltheory}
we describe how invariant DODS are constructed for given symmetry
algebras. We also show how symmetries can be used to find particular
(namely, invariant) solutions of DODSs. Section~\ref{linearsec} is
devoted to invariant linear DODSs and their invariant solutions. In
Section~\ref{mostsymmetric} we consider a special DODE, which
appears often in the classification. 
Finally, the conclusions are presented in
Section~\ref{Conclusions} 
and the results are summed
up in Table~2.

\section{Construction of invariant linear DODS and their invariant solutions}

\label{generaltheory}

In  Section~\ref{linearsec}  below we shall run through the list of Lie algebras of vector fields~\cite{ourpaper} in
two variables, concentrating on the Lie algebras excluded in~\cite{ourpaper}. 
They are the Lie algebras containing
algebras~(\ref{Dcase24}) and~(\ref{Dcase22}) as {sub}algebras. 
The invariant DODS thus obtained is linear as in~(\ref{DODEcase24}) or~(\ref{DODEcase22}).

The invariant DODS are obtained in the form
\begin{equation}   \label{implicit1}
F_1 (   I_1, ... , I_k ) = 0  ,   \quad    F_2 (   I_1, ... , I_k ) = 0   ,
\end{equation}
$$
\mbox{det}  \  {\partial ( F_1 , F_2 )  \over \partial ( \dot{y}  , x_- ) } \neq 0,
$$
where $ I_1, ... , I_k $ are invariants of the considered Lie group.
These can be "strong invariants"
$$
\mbox{\bf pr} X  _{\alpha}   F_a ( x, y ,x_- , y_- ,\dot{y} ) = 0 ,
\qquad
 \alpha = 1, ... , n  , \quad  a=1,2,
$$
or "weak invariants" satisfying
\begin{equation}
\left. 
\mbox{\bf pr} X  _{\alpha}   F_a ( x, y ,x_- , y_- ,\dot{y} )
\right|_{F_1=F_2=0} = 0 ,
\qquad
 \alpha = 1, ... , n  , \quad  a=1,2 . 
\end{equation}
Weak invariants are only invariant on the manifold $F_1=F_2=0$, 
i.e. on the solutions of the linear DODS~(\ref{implicit1}).

The invariant equation~(\ref{implicit1}) can always be rewritten in
the form~(\ref{DODEcase24}) or~(\ref{DODEcase22}). Once the
equations are obtained we search for exact solutions, so called
group invariant solutions, using symmetry reduction.

In the case of the DODS~(\ref{DODE})
it is sufficient to consider one-dimensional {sub}algebras.
All of them have the form
\begin{equation}    \label{alloperators}
X = \sum  _{ \alpha = 1 } ^n   c _{ \alpha}   X _{ \alpha} ,
 \qquad
  c _{ \alpha}  \in \mathbb{R}  ,
\end{equation}
where   $c _{ \alpha} $ are constants and $  X _{ \alpha}  $ are of the form~(\ref{operator1}) 
and are elements of the symmetry algebra $L$ of the considered DODS.

\bigskip

The method consists of several steps.

\begin{enumerate}

\item

Construct a representative list of one-dimensional {sub}algebras of   $L_i $
of the symmetry algebra $L$ of the DODS.
The  {sub}algebras are classified under the group of inner automorphisms $ G  = \mbox{exp} L$.

\item

For each {sub}algebra in the list calculate the invariants of the
subgroups    $ G_i   = \mbox{exp} L_i $ in the four-dimensional
space with local coordinates $ \{  x, y, x_- , y_- \} $. There are
three functionally independent invariants. For the method to be
applicable two of the invariants must depend on two variables only,
namely  $(x,y) $ and   $(x_- ,y_- ) $ respectively.
\par
We denote 3
invariants $J_1 (x,y)$, $J_2 (x_-,y_-)$ and $J_3  ( x, y, x_- , y_- ) $. 
We put two of them equal to constants, namely:
\begin{equation}   \label{invariantsform}
J_1 (x,y) =A,  \qquad  J_3  ( x, y, x_- , y_- )  = B .
\end{equation}
They  must satisfy the Jacobian condition
\begin{equation}
\mbox{det}   \left(  { \partial ( J_1 , J_3 )  \over  \partial ( y
, x _- ) }  \right)     \neq   0
\end{equation}
 (we have  $ J_2  (  x_- , y_- )  =   J_1  ( x_- , y_- )=A  $,
i.e.,    $ J_2  (  x_- , y_- )  $ is obtained by shifting  $ J_1  (
x, y)  $ to $ (  x_- , y_- ) $).

All elements of the Lie algebra have the form~(\ref{alloperators}).
A {\bf necessary condition}  for invariants of the form~(\ref{invariantsform}) to exist
is that at least one of the vector fields in $L_i$  satisfies
\begin{equation}  \label{reduction_condition}
 \xi (x,y) {\not \equiv} 0   .
\end{equation}   
Only vector fields in the representative list satisfying~(\ref{reduction_condition}) 
will provide invariant solutions.

\item

Solve equations~(\ref{invariantsform})  to obtain the reduction formulas
\begin{equation}  \label{reduction_formulas}
y = h (x, A) ,
\qquad
x_-  = k ( x, A, B)
\end{equation}
(we also have $y_-  = h (x_- , A)$).

\item

Substitute the reduction formulas~(\ref{reduction_formulas})
into the DODS~(\ref{DODE})   and require that the equations are satisfied  identically.
This 
provides relations which define the constants
and therefore determine the functions $h$  and $k$.
It may also impose constraints on the functions $ f ( x, y, y_-)$  and $ g ( x, y, y_-)$
in~(\ref{DODE}) 
and on the values of parameters that may figure in the {sub}algebras 
used in the reduction. 
Once the relations are satisfied the invariant solution is given by~(\ref{reduction_formulas}).

\item

Additional step, which is not implemented in the present paper.
Apply the entire group $G_i = \mbox{exp} L_i $ to the obtained invariant solutions.
This can provide a more general solution depending on up to three more parameters
(corresponding to the factor algebra $ L / L_i$).
Check whether the DODS imposes further constraints involving the new-parameters.

\end{enumerate}


\section{Invariant DODSs and invariant solutions}

\label{linearsec}

In this section we 
rely on realizations of three- and four-dimensional Lie algebras
by real vector fields of the form~(\ref{operator1}),
which are used in~\cite{ourpaper} and introduced in~\cite{Gonzalez1992}.
They are given in Table~1.
For the present paper we select only realizations which contain two-dimensional {sub}algebras of  linearly connected vector fields.
Complete tables of all algebras can be found in~\cite{ourpaper}.

We  run through the list of representative algebras
and for each of them construct all group invariants in the jet space with local coordinates
$ \{ x  , y , x_- , y_- , \dot{y}  \}  $.
We  construct both strong invariants (invariant  in the entire space)
and weak invariants (invariant on some manifolds).
The invariants will be used to write invariant DODSs.
In some cases we 
use the notations
$$
\Delta x = x - x_- ,
\qquad
\Delta y = y - y_- .
$$

We 
also look for invariant solutions as described in the preceding  section.
Representative optimal systems of one-dimensional {sub}algebras are taken from~\cite{PateraWinternitz}.

\subsection{Dimension 3}

\label{dim3}

\bigskip

$\mbox{\bf A}_{3,1}$:
The nilpotent Lie algebra   $ {\sso {n}} _{3,1} $ can be realized as
\begin{equation}       \label{symmetry_A31}
X_1 = { \ddy} ,
\qquad
X_2 = x { \ddy} ,
\qquad
X_3 =   { \ddx}  .
\end{equation}
A basis of the invariants is given by
$$
I_1 = \Delta x ,
\qquad
I_2 =  \dot{y}  -    { \Delta y  \over \Delta x  }   .
$$
It provides us with an invariant DODS
\begin{equation}      \label{DODS_31}
 \dot{y}  =    { \Delta y  \over \Delta x  }   +  C_1  ,
\qquad
\Delta x  = C_2   ,
\qquad
  C_2  > 0 .
\end{equation}

We now show how to find invariant solutions following the procedure
given in Section 2:

\medskip

Step 1. The representative algebra of one-dimensional {sub}algebras (see~\cite{PateraWinternitz}) is
$$
\{  X_1 \} ,
\qquad
\{  \cos ( \varphi ) X_2 + \sin ( \varphi ) X_3 \}    .
$$
Here and below $    0 \leq \varphi  < \pi $.
Due to the condition $ \xi (x,y) {\not \equiv} 0 $
invariant solutions  can exist only for the second element with $  \sin ( \varphi ) \neq 0$.
In this case we can rewrite this operator as $ a X_2 + X_3 $.
Here and in the following cases $ - \infty < a < \infty $
unless another interval is given.

Step 2. The invariants for $ a X_2 + X_3 $ 
in the space $ \{ x  , y , x_- , y_-  \}  $ are
\begin{equation*}    \label{invariants_A31}
J_1 = y  -  { a \over 2 } x^2  = A ,
\qquad
J_3 = { x   - x_-   } = B .
\end{equation*}

Step 3. The reduction formulas take the form
\begin{equation}        \label{solution_form_A31}
y =   { a \over 2 } x^2  +  A  ,
\qquad
x_- = x - B   .
\end{equation}

Step 4. The substitution in~(\ref{DODS_31}) gives restrictions
\begin{equation*}        \label{restrictions_A31}
{ a \over 2 } B = C_1  ,
\qquad
B = C_2  \neq 0 .
\end{equation*}
Thus, for $ a = 2 { C_1 \over C_2 } $ there is the invariant solution~(\ref{solution_form_A31})
with arbitrary $A$ and $ B=  C_2 $. 
The invariant DODS~(\ref{DODS_31}) depends on two constants $C_1$ and $C_2$. 
The solution~(\ref{solution_form_A31}) depends on three constants. 
One of them ($A$) is free, the others  $a$ and $B$ are expressed in terms of $C_1$ and $C_2$.

\medskip

For the following cases we 
employ  the same steps while providing less details.

\bigskip

$\mbox{\bf A}_{3,3} ^a $:
The algebra $ {\sso {s}} _{3,1} $ has two realizations~\cite{ourpaper}.
One of then, namely
\begin{equation}          \label{symmetry_A33}
X_1 = { \ddy} ,
\qquad
X_2 = x { \ddy},
\qquad
X_3 =   ( 1 - a ) x  { \ddx}   +   y { \ddy} ,
\qquad
 0 < |a| \leq 1 ,
\end{equation}
contains linearly connected vector fields (the {nil}radical $ \{  X_1, X_2 \} $).

We  consider two {sub}cases:

\begin{itemize}

\item

 $ a \neq 1 $

A basis of invariants is
$$
I_1 =      x^{a \over a-1}   \left(  \dot{y}  -    { \Delta y  \over \Delta x  }  \right)  ,
\qquad
I_2 = { x_-  \over x}
$$
and the general invariant DODS can be written as
\begin{equation}      \label{DODS_A33a}
 \dot{y}  =    { \Delta y  \over \Delta x  }   +  C_1   x^{a \over 1-a}  ,
\qquad
 x_-     =  C_2  x   ,
\qquad
( 1 - C_2  ) x > 0  .
\end{equation}

The representative list of one-dimensional {sub}algebras  has four elements:
$$
\{  X_1 \}   ,
\qquad
\{  X_2 \}  ,
\qquad
\{  X_1 \pm X_2 \}    ,
\qquad
\{  X_3 \}  .
$$
Invariant solutions exist only for the last element.

For $ X_3 $ we find invariant solutions in the form
\begin{equation}        \label{solution_form_A33a}
y =   A    x^{ 1 \over 1 - a }    ,
\qquad
x_-  = B  x   ,
\end{equation}
with constants $A$ and $B$ given by the system
\begin{equation*}      \label{restrictions_A33a}
A \left( { 1 \over 1 - a  }  -    { 1 - {C_2} ^{ 1 \over 1 - a }   \over 1 - C_2 }  \right)  =     C_1    ,
\qquad
B  =  C_2   .
\end{equation*}

\item

$ a = 1 $

There are two invariants
$$
I_1 =  x ,
\qquad
I_2 = x_-
$$
and invariant manifold
$$
   \dot{y}  -    { \Delta y  \over \Delta x  }   = 0   .
$$
We obtain the invariant DODS
\begin{equation}      \label{DODS_A33b}
 \dot{y}  =   { \Delta y  \over \Delta x  } ,
\qquad
 x_-   = g (x)     ,
\qquad
   g (x) < x ,
\qquad
 g(x) {\not\equiv}  \mbox{const}    .
\end{equation}
In this case there are no invariant solutions
because there are no symmetries satisfying $ \xi (x,y) {\not \equiv} 0 $.


\end{itemize}

\bigskip

$\mbox{\bf A}_{3,5}$:
The solvable algebra  ${\sso {s}} _{3,2}$ has two realizations and one of them,
namely
\begin{equation}          \label{symmetry_A35}
X_1 = { \ddy} ,
\qquad
X_2 = x { \ddy} ,
\qquad
X_3 =  { \ddx}  +  y { \ddy}
\end{equation}
contains the subalgebra $\mbox{\bf A}_{2,3}$.
A basis of invariants is
$$
I_1 =    \Delta x ,
\qquad
I_2 =    e^{-x}  \left(  \dot{y}    -   { \Delta y  \over \Delta x  }  \right)    .
$$
We find the general invariant DODS
\begin{equation}        \label{DODS_A35}
 \dot{y}  =    { \Delta y  \over \Delta x  }   +  C_1    e^x ,
\qquad
\Delta x  = C_2    ,
\qquad
C_2 > 0   .
\end{equation}

The optimal system of one-dimensional {sub}algebras is
$$
\{ X_1  \}  ,
\qquad
\{  X_2 \}  ,
\qquad
\{  X_3 \}
$$
and invariant solutions exist only for  $ \{  X_3 \}  $.
We obtain  the solution
\begin{equation}        \label{solution_form_A35}
y =   A    e^x     ,
\qquad
x_-  = x - B   ,
\end{equation}
for constants   $A$ and $B$ satisfying the system
$$
A = \left({ C_1C_2 } \over  {C_2-1 +  e ^{ - C_2 }  } \right) ,
\qquad  B  =  C_2     .
$$

\bigskip

$\mbox{\bf A}_{3,7}^b$:
There are two realizations of the algebra  $ {\sso {s}} _{3,3} $,
and one of them contains the subalgebra $\mbox{\bf A}_{2,3}$.
It is
\begin{equation}          \label{symmetry_A37}
X_1 = { \ddy} ,
\qquad
X_2 = x { \ddy} ,
\qquad
X_3 =   ( 1   +  x ^2 )  { \ddx}   +  ( x + b)   y { \ddy} ,
\quad
b \geq 0 .
\end{equation}
A basis of the invariants
$$
I_1 = {  x - x_- \over 1 + x x_- } ,
\qquad
I_2 =   \sqrt{ 1 + x^2 }     e^{ - b   \arctan (x) }  \left(   \dot{y}  -    { \Delta y  \over \Delta x  }  \right) 
$$
provides us with the most general invariant DODS
\begin{equation}     \label{DODS_A37}
 \dot{y}  =    { \Delta y  \over \Delta x  }   +  C_1   { e^{ b   \arctan (x) }  \over  \sqrt{ 1 + x^2 }  } ,
\qquad
x_- = { x   - C_2   \over   1 +  C_2  x }    ,
\qquad
{  C_2   \over   1 +  C_2  x }  > 0 .
\end{equation}

The representative list of {sub}algebras is
$$
\{ X_2 \} ,
\qquad
\{ X_3 \}  ;
$$
invariant solutions exist only for $ X_3 $.
We obtain solutions of the form
\begin{equation}        \label{solution_form_A37}
y =   A    \sqrt{ 1 + x^2 } e^{ b \arctan x }    ,
\qquad
x_-  = { x - B  \over 1 + B x }
\end{equation}
with constants $A$ and $B$ defined by the system
$$
A \left(     b - { 1  \over  C_2  }
+   { \sqrt{  1+  C_2 ^2 } \over  C_2  }   e^{ -  b \arctan  C_2  }  \mbox{sign} ( 1 +  C_2  x )      \right)
=   C_1    ,
\qquad
B  =  C_2     .
$$

\bigskip

$\mbox{\bf A}_{3,11}$:
There are four realizations of algebra $ {\sso {sl}}  (2, \mathbb{R}) $.
Only one of them contains linearly connected vector fields.
It can be presented as
\begin{equation}          \label{symmetry_A311}
X_1 =   { \ddy} ,
\qquad
X_3 = y  { \ddy} ,
\qquad
X_3 = y^2   { \ddy}   .
\end{equation}
Note that all operators are linearly connected.
There are two invariant
$$
I_1 = x,
\qquad
I_2 =x_-
$$
and two invariant manifolds
$$
y - y_- = 0
\qquad
\mbox{and}
\qquad
\dot{y} = 0 .
$$
In this case there is no invariant DODS.


\bigskip

$\mbox{\bf A}_{3,13}$:
There exist two realizations of the decomposable algebra  $ {\sso {n}}_{1,1}    \oplus {\sso {s}}_{2,1} $
and both of them contain linearly connected vector fields.
The first realization can be taken as
 \begin{equation}          \label{symmetry_A313}
X_1 = { \ddx} ,
\qquad
X_2 = { \ddy} ,
\qquad
X_3 =  y { \ddy}   .
\end{equation}
We find the invariants
$$
I_1 =   \Delta x ,
\qquad
I_2 =   { \Delta x  \over \Delta y  } \dot{y}
$$
and obtain the invariant DODS
\begin{equation}      \label{DODS_A313}
 \dot{y}  =    C _1 { \Delta y  \over \Delta x  } ,
\qquad
\Delta x  =  C_2   ,
\qquad
C_2 > 0 .
\end{equation}

The optimal system
$$
\{  X_2  \} ,
\qquad
 \{  X_1 \pm  X_2 \} ,
\qquad
\{  \sin ( \varphi )  X_1    +  \cos ( \varphi )  X_3   \}
$$
contains two elements for which we can find invariant solutions:

\begin{itemize}

\item

For $  X_1 \pm  X_2 $ we find invariant  solutions
\begin{equation}        \label{solution_form_A313_a}
y = \pm  x +   A       ,
\qquad
x_-  = x - B     ,
\end{equation}
only if $  C_1 = 1 $.
In this case $ A  $ is arbitrary and $ B  =  C_2   $.

\item

For $ X_1 + a  X_3 $
(it is  $  \sin ( \varphi )  X_1    +  \cos ( \varphi )  X_3    $ with  $ \sin ( \varphi ) \neq  0 $)
we find the solution
\begin{equation}        \label{solution_form_A313_b}
y =   A    e ^{  a x }     ,
\qquad
x_-  = x - B     ,
\end{equation}
with constants $A$ and $B$ satisfying
$$
  A  a
= C_1  A  {   1 - e^{ - a C_2  } \over C_2  }   ,
\qquad
B  =  C_2   .
$$
If $a$ satisfies
$$
a   = C_1   {   1 - e^{ - a C_2 } \over C_2   },
$$
we can take arbitrary $A$ and $ B =C_2 $.
If not, we get only the trivial solution $ y = 0 $, $ x_- = x - C_2$.

\end{itemize}

\bigskip

$\mbox{\bf A}_{3,14}$:
The second realization of the algebra $ {\sso {n}}_{1,1}    \oplus {\sso {s}}_{2,1} $ is
\begin{equation}          \label{symmetry_A314}
X_1 = x { \ddy} ,
\qquad
X_2 = { \ddy},
\qquad
X_3 =    x  { \ddx}   +   y { \ddy}   .
\end{equation}
We obtain the invariants
$$
I_1 = {  x_-   \over x  }   ,
\qquad
I_2 = \dot{y}  -    { \Delta y  \over \Delta x  }
$$
and invariant DODS
\begin{equation}      \label{DODS_A314}
 \dot{y}  =    { \Delta y  \over \Delta x  }   +  C_1  ,
\qquad
 x_-     =  C_2  x   ,
\qquad
( 1- C_2 ) x > 0  .
\end{equation}

The representative list of one-dimensional {sub}algebras is
$$
\{  X_2  \} ,
\qquad
 \{  X_1 \pm  X_2 \} ,
\qquad
\{  \sin ( \varphi )  X_1    +  \cos ( \varphi )  X_3   \}   .
$$
There can be invariant solutions only for
$  \sin ( \varphi )  X_1    +  \cos ( \varphi )  X_3     $ with  $ \cos ( \varphi )   \neq 0 $.
We rewrite these operators as  $ a X_1 + X_3 $.
The invariant solutions have the form
\begin{equation}        \label{solution_form_A314}
y =   a x \ln |x|  + A x     ,
\qquad
x_-  = B x     ,
\end{equation}
with 
\begin{equation}        \label{solution_form_A314_constants}
a  \left( 1 -   { C_2   \ln |C_2 |    \over   1 - C_2   } \right)  =   C_1   ,
\qquad
B  =  C_2   .
\end{equation}       
Thus, the invariant solutions exist only for $a$   satisfying~(\ref{solution_form_A314_constants}). 
In this case $A$ is arbitrary and  $ B  =  C_2   $.

\bigskip

$\mbox{\bf A}_{3,15}$:
For the Abelian algebra $ 3{\sso {n}}_{1,1} $
we get the following realization
\begin{equation}          \label{symmetry_A315}
X_1 =   { \ddy}  ,
\qquad
X_2 = x   { \ddy} ,
\qquad
X_3 =  \chi (x)   { \ddy}  ,
\qquad
 \ddot{\chi}   (x)  \neq  0    .
\end{equation}
We remark that all operators are  linearly connected.

There are two invariants
$$
I_1 = x ,
\qquad
I_2 = x_-
$$
and invariant manifold
$$
  \dot{y} - { \Delta y \over \Delta x  }  = 0
$$
provided that
\begin{equation}   \label{symmetry_A178}
\dot{\chi} (x)   = { \chi (x) - \chi ( x_-)  \over  x  - x_-   }     .
\end{equation}
Thus, we can present the most general invariant DODS as
$$
  \dot{y} =  { \Delta y \over \Delta x  } = f( x) ,
\qquad
 \dot{\chi} (x)   = { \chi (x) - \chi ( x_-)  \over  x  - x_-   }  .
$$
where  $ f( x) $ is an arbitrary function.
Note that the second equation is the delay relation which defines $x_-$
as an implicit function of $ x $.

The symmetries~(\ref{symmetry_A315}) provide no invariant solutions.

\bigskip

To sum up, for three-dimensional Lie algebras containing ${\bf A}_{2,1}$ or ${\bf A}_{2,3}$ 
as {sub}algebras, we have presented all corresponding DODS and each class is represented by a linear DODS. 
For ${\bf A}_{3,1}$, ${\bf A}_{3,3} ^a $ with $ a\neq 1$,  ${\bf A}_{3,5}$, ${\bf A}_{3,7} ^b $, 
${\bf A}_{3,13}$ and  ${\bf A}_{3,14}$ their DODS depend on 2 or 3 parameters. 
${\bf A}_{3,3} ^a $ with $ a = 1 $  provides the DODS~(\ref{DODS_A33b}) 
which depends on one function $ g (x) $. 
For ${\bf A}_{3,15}$ the DODS involves a function $ \chi (x) $ 
satisfying the linear DODS ~(\ref{symmetry_A178}).
All obtained  invariant solutions are quite explicit and correspond to very specific invariant conditions. The constants in the solutions are expressed in
terms of those in the DODS, 
or may take arbitrary values.


\subsection{Dimension 4}

\label{dim4}

In this section we go through realizations of four-dimensional Lie algebras
given in Table~1. For most of these realizations there are no invariant DODSs. 
We present only the four realizations which provide us with invariant DODSs.

\bigskip

$\mbox{\bf A}_{4,5}$:
One of the realizations of  the solvable algebra $ {\sso {s}}_{4,3} $ is given
by the operators
\begin{equation}          \label{symmetry_A45}
X_1 =   { \ddy}  ,
\qquad
X_2 = x   { \ddy} ,
\qquad
X_3 =  \chi (x)   { \ddy}  ,
\qquad
X_4 = y   { \ddy} ,
\qquad
 \ddot{\chi}   (x)  \neq  0    .
\end{equation}
All four vector fields are linearly connected. 
There are two invariants
$$
I_1 = x ,
\qquad
I_2 = x_-
$$
and invariant manifold
$$
  \dot{y} -  { \Delta y \over \Delta x  }  =  0
$$
provided that
\begin{equation}   \label{DELAYcase416}
 \dot{\chi} (x)   = { \chi (x) - \chi ( x_-)  \over  x  - x_-   }   .
\end{equation}
Thus we get the invariant DODS
\begin{equation}    \label{DODS_A45}
  \dot{y} =  { \Delta y \over \Delta x  } ,
\qquad
 \dot{\chi} (x)   = { \chi (x) - \chi ( x_-)  \over  x  - x_-   }    ,
\end{equation}
where the last equation is the delay relation,
which defines $ x_- $ as a function of $ x $.

There are no invariant solutions for  symmetries~(\ref{symmetry_A45})
since they do not satisfy the condition~(\ref{reduction_condition}).


\bigskip

$\mbox{\bf A}_{4,12}$:
One of two realization of algebra $ {\sso {s}}_{4,11} $
is given by the operators
\begin{equation}         \label{symmetry_A412}
X_1 =  { \ddx} , 
\qquad
X_2 =  x { \ddy} ,
\qquad
X_3 =   { \ddy}  ,
\qquad
X_4 =  y  { \ddy} .
\end{equation}
There is one invariant
$$
I = \Delta x
$$
and one invariant manifold
$$
 \dot{y}   -  {  \Delta   y   \over \Delta x }  = 0    .
$$
We obtain the most general invariant DODS
\begin{equation}    \label{DODS_A412}
 \dot{y}   =  {  \Delta   y   \over \Delta x }  ,
\qquad
  \Delta  x     =  C    ,
\qquad
C > 0 .
\end{equation}

The representative list of one-dimensional {sub}algebras
consists of five algebras 
$$
\{ X_1   \}   ,
\qquad
\{ X_2   \}  ,
\qquad
\{  X_1 \pm  X_2   \} ,
\qquad
\{ X_3  \} ,
\qquad
\{  a X_1 + X_4   \}   .
$$
Invariant solution can be found for the following three:

\begin{itemize}

\item
Invariant solution for $ X_1  $  is
\begin{equation}        \label{solution_form_A412_a}
y =   A     ,
\qquad
x_-  = x - B
\end{equation}
with arbitrary $A$ and $ B  =  C $.

\item

For $  X_1 \pm X_2  $
the invariant solutions have the form
\begin{equation}        \label{solution_form_A412_b}
y =   \pm { x^2 \over 2 } + A     ,
\qquad
x_-  = x - B     .
\end{equation}
In this case invariant solutions do not exist because the obtained conditions
$ B =0 $ and $ B  =  C $ are not compatible with $ C > 0 $.

\item
For $ a X_1 + X_4 $ (only for $ a \neq 0$)
invariant solutions have the form
\begin{equation}        \label{solution_form_A412_c}
y =   A e ^{ x / a }      ,
\qquad
x_-  = x - B       
\end{equation}
with the restrictions
$$
A \left(    { 1 \over a }   -  { 1 - e ^{ - C a}   \over   C   }    \right)  =  0  ,
\qquad
B  =  C   .
$$

If the constant $a$ satisfies
$$
{1 \over a }   = { 1 - e ^{ - C a}   \over   C },
$$
then $A$ is arbitrary and $ B  =  C $ in~(\ref{solution_form_A412_c}). 
Otherwise we only get the trivial solution
$$
y =   0       ,
\qquad
x_-  = x - C     .
$$

\end{itemize}

\bigskip

$\mbox{\bf A}_{4,14}$:
Algebra   $  {\sso {s}}_{4,12}  $ has a realization
\begin{equation}         \label{symmetry_A414}
X_1 =   { \ddy} ,
\qquad
X_2 = x  { \ddy} ,
\qquad
X_3 =  y   { \ddy} ,
\qquad
X_4 = ( 1+ x^2 )  { \ddx}  +   x y  { \ddy}  .
\end{equation}
We find one invariant
$$
I =  { x - x_- \over 1 + x x_- }
$$
and invariant manifold
$$
 \dot{y}   -  {  \Delta   y   \over \Delta x }  = 0   .
$$
We obtain the most general invariant DODS
\begin{equation}    \label{DODS_A414}
\dot{y}   =  {  \Delta   y   \over \Delta x } ,
\qquad
x_- = { x   - C   \over   1 +    C  x }    ,
\qquad
{ C   \over   1 +    C  x }  > 0 .
\end{equation}

The optimal system of {sub}algebras contains three elements
$$
\{  X_1  \}   ,
\qquad
\{  X_3  \} ,
\qquad
\{ a X_3 + X_4 \}
$$
and invariant solutions exist only for the  element $ a X_3 + X_4  $.
They have the form
\begin{equation}        \label{solution_form_A414}
y =   A \sqrt{ 1 + x^2 } e ^{ a  \arctan x }      ,
\qquad
x_-  = { x - B  \over 1 + B x }
\end{equation}
provided that $a$, $ A$   and $B$ satisfy restrictions
$$
A \left( a -  { 1 \over C  }
+  { \sqrt{ 1 + C^2 } \over C }  e ^{ - a  \arctan C  }    \mbox{sign} ( 1 + C x )    \right) =  0    ,
\qquad
B  =  C  .
$$
If $ a$  and $ C $ satisfy
$$
 a -  { 1 \over C  }
+  { \sqrt{ 1 + C^2 } \over C }  e ^{ - a  \arctan C  }    \mbox{sign} ( 1 + C x )  =  0   ,
$$
we obtain invariant solution~(\ref{solution_form_A414}) with arbitrary $ A $  and  $ B = C $.
If not, we get only trivial invariant solution $ A = 0$ (i.e., $ y = 0 $) and $ B = C $.

\bigskip

$\mbox{\bf A}_{4,21}$:
For the decomposable algebra   $ 2 {\sso {s}}_{2,1}   $
there is a realization
\begin{equation}          \label{symmetry_A421}
X_1 =   { \ddy} ,
\qquad
X_2 =  x  { \ddx}  +  y   { \ddy}  ,
\qquad
X_3 =  x  { \ddx} ,
\qquad
X_4 =  x   { \ddy}  .
\end{equation}
We find one invariant
$$
I =  { x_- \over x }
$$
and the invariant manifold
$$
 \dot{y}   -  {  \Delta   y   \over \Delta x }  = 0  .
$$
They provide us with  the most general invariant DODS
\begin{equation}    \label{DODS_A421}
 \dot{y}   =  {  \Delta   y   \over \Delta x }   ,
\qquad
  x _-       = C x ,
\qquad
( 1 - C )  x > 0 .
\end{equation}

For convenience  we chose a simpler basis for the realization~(\ref{symmetry_A421}) as
\begin{equation}          \label{DDcase42b}
Y_1 =   x { \ddx} ,
\qquad
Y_2 =  { \ddy} ,
\qquad
Y_3 =  x   { \ddy} ,
\qquad
Y_4 =  y  { \ddy} .
\end{equation}
Then we obtain the following optimal system of one-dimensional sub{algebras}
$$
\{   Y_1   \}   ,
\qquad
\{   Y_2    \} ,
\qquad
\{   Y_1 \pm  Y_2    \} ,
\qquad
\{   Y_3   \} ,
\qquad
\{   Y_2 \pm  Y_3   \} ,
\qquad
\{   a Y_1 + Y_4  \}   .
$$
We find the following invariant solutions:

\begin{itemize}

\item

Invariant solutions for  $  Y_1 $ have form
\begin{equation}        \label{solution_form_A421_a}
y =   A       ,
\qquad
x_-  = B  x     ,
\end{equation}
with arbitrary $A$ and  $  B  =  C $.

\item

For $ Y_1 \pm Y_2  $ we look for solutions
\begin{equation}        \label{solution_form_A421_b}
y =   \pm \ln  | x |  +  A       ,
\qquad
x_-  = B  x
\end{equation}
and get conditions
$$
  {  \ln | C |   \over C  - 1 }   =  1 ,
\qquad
B  =  C    .
$$
The invariant solutions exist only if   $  {  \ln | C |  = C - 1  } $.
In this case $ A $ can be arbitrary.

\item

For $ a Y_1 + Y_4  $ (only for $ a \neq 0$)
we look for invariant solutions in the form
\begin{equation}        \label{solution_form_A421_c}
y =   A  x ^{ 1  / a  }      ,
\qquad
x_-  = B  x
\end{equation}
and get restrictions
$$
A \left( { 1 \over a  }  -   {  1 - C   ^{ 1  / a  }  \over 1 - C  }  \right) = 0   ,
\qquad
B  =  C   .
$$
If $ a$ and $ C $ satisfy the relation
$$
{ 1 \over a  }  -   {  1 - C   ^{ 1  / a  }  \over 1 - C  }   = 0   ,
$$
we obtain invariant solutions~(\ref{solution_form_A421_c})
with arbitrary $ A $ and $ B = C $.
If not, we get only trivial invariant solution $  y = 0 $, $ x_- = C x $.

\end{itemize}

\subsection{Higher dimensional cases}

\label{dimH}

Direct computation shows that
for realizations of  Lie algebras of dimension $ n \geq 5 $
by vector fields in a plane
we do not get new cases of DODSs.
However, some cases that are determined by Lie algebras of dimension $ 2 \leq n \leq  4 $
can also be invariant under higher dimensional finite Lie algebras.
This happens because linear DODEs with solution independent delay relations
admit infinite-dimensional symmetry groups 
that may contain finite-dimensional {sub}algebras of higher dimension.

For example, the DODS~(\ref{DODS_A412})
admits infinitely many symmetries of the form
\begin{equation}    \label{exp_symmetry}
X = e^{ \lambda  x} {\ddy} ,
\end{equation}
where $\lambda$ is a solution of the characteristic equation
\begin{equation}    \label{character}
\lambda = { 1 -  e^{ - \lambda  C }  \over C }   .
\end{equation}
This equation can be rewritten as
\begin{equation}    \label{character2}
 1 + z = e ^z
\end{equation}
for $z = - \lambda C $.
Except for the operator $ \partial _ y $ ($\lambda = 0$)
the symmetries~(\ref{exp_symmetry}) are complex, but we can obtain real symmetries
taking real and imaginary parts of these complex  symmetries.
Some subsets of such symmetries together with symmetries~(\ref{symmetry_A412})
are higher dimensional realizations of finite-dimensional Lie algebras.
Let us consider a particular example.

\begin{example} \label{differentgroups}

Equations~(\ref{DODS_A412})
are  specified by the four-dimensional symmetry algebra~(\ref{symmetry_A412}).
The same equations also allow the six-dimensional symmetry algebra 
$$
X_1 =  { \ddx} ,
\qquad
X_2 =  { \ddy} ,
\qquad
X_3 =  x  { \ddy}  ,
\qquad
X_4 =  y  { \ddy} ,
$$
$$
X_5 = e ^{ a x } \cos(bx)   { \ddy} ,
\qquad
X_6 =   e ^{ a x } \sin(bx)  { \ddy}  ,
$$
where $ \lambda = a + i b \neq 0 $ is a solution of~(\ref{character}).
\end{example}

Since we do not get new DODSs for Lie algebras of dimension $ n \geq 5 $
we conclude that we have listed all linear DODSs which can be specified
by finite-dimensional Lie symmetry algebras. 
The larger symmetry algebras may however provide new invariant
solutions. 
For instance, in Example~\ref{differentgroups} 
one can find particular solutions which are invariant for the {sub}algebras $X_1 + \lambda X_5$.

\begin{remark}
It is interesting to note that Eq.~(\ref{character2}) can be represented
with the help of Bernoulli numbers (see~\cite{Gelfond, Jordan}).
$$
{ z \over  1 -  e^{-z}  }  = \sum _{n = 0} ^{\infty} { B_n \over n! } ( -z) ^n =1
$$

\end{remark}

\section{Comments on a  special case of a linear DODE}

\label{mostsymmetric}

In this section we show how integration procedure is applied to a particular DODS.
We consider delay differential equation
\begin{equation}      \label{most}
 \dot{y} = { \Delta  y   \over \Delta x  }
\end{equation}
for the delay relation
\begin{equation}      \label{mostmesh}
{x}_-  = q x - \tau  ,
\end{equation}
where  $  q > 0 $ and $ \tau \geq 0 $  are constants such that
$   \tau ^2 + (q-1)^2 \neq 0 $.
This delay relation can be considered as a generalization of constant delay $ \Delta x  = \tau  $ (for $q=1$)
and $q$-delay relation ${x}_-  = q x $ (for $\tau = 0$).
Note  that relation~(\ref{mostmesh})
is equivalent to delay relation
\begin{equation}
   \Delta x    = ( 1 - q ) x + \tau  .
\end{equation}

\subsection{Symmetries}

The DODS admits symmetries
\begin{equation}
X_1  =  (  (q - 1)   x  - \tau   )  { \ddx} ,
\qquad
X_2  =  y  { \ddy} ,
\qquad
X_{ \alpha}  = \alpha (x)  { \ddy} ,
\end{equation}
where $\alpha (x)$ is an arbitrary solution of the DODE~(\ref{most})
with delay relation~(\ref{mostmesh}).
Generally,  solutions  $\alpha (x)$  are piecewise-smooth.

\begin{remark}
This symmetry algebra includes a {sub}algebra with smooth coefficients given by the operators
\begin{equation}    \label{finite}
X_1  =  (  (q - 1)   x  - \tau    )  { \ddx} ,
\qquad
X_2 = y { \ddy} ,
\qquad
X_3 =  { \ddy} ,
\qquad
X_4 = x { \ddy} .
\end{equation}
If we require invariance with respect to these symmetries,
we obtain the DODS~(\ref{most}),(\ref{mostmesh}).

Operators~(\ref{finite})  are equivalent 
to  operators~(\ref{symmetry_A412}) 
representing algebra  $\mbox{\bf A}_{4,12} $ 
if $q = 1 $ and $\tau \neq  0$ 
and 
to  operators~(\ref{symmetry_A421})   
representing algebra  $\mbox{\bf A}_{4,21}$ 
if $ q \neq 1 $ 
(in the first case we can simply divide $ X_1 $ by $\tau $, 
in the second we can divide $X_1$ by $q-1$ and then  
remove $\tau$ by the translation  $ x \rightarrow   x - \tau / (q-1)$).  
Therefore, one can transform invariant solutions of cases  
$\mbox{\bf A}_{4,12}$ and $\mbox{\bf A}_{4,21}$ 
into invariant solutions of DODS~(\ref{most}),(\ref{mostmesh}).



\end{remark}

\subsection{Sequences of intervals}

Consider several cases of delay relation~(\ref{mostmesh}) in detail.
Solving Eq.~(\ref{mostmesh})
for the initial interval $[ x_{-1} , x_0 ] $,
such that $ {x}_{-1}  =  q x_0 - \tau $,
we can obtain the sequence of intervals  explicitly.

\begin{enumerate}

\item

If $q=1 $,  we consider the delay relation
\begin{equation}
\Delta x  =  \tau ,
\end{equation}
which is invariant with respect to the symmetry
$$
X_1  = { \ddx}   .
$$
We obtain the points
\begin{equation}  \label{uniform}
x_n = x_0 + n \tau ,
\qquad
\tau = x_0 - x_{-1},
\qquad
n = -1 , 0, 1, 2, ...
\end{equation}

\item

If $q \neq 1 $,  we obtain the points
\begin{multline}  \label{qmesh}
x_n
=  { x_0  \over 1 - q }   \left( { 1 \over q^n } - q \right)
+  { x_{-1}   \over  q - 1  }   \left( { 1 \over q^n } - 1 \right)
\\
= { x_0    \over q^n }
+  { \tau   \over  1 - q  }   \left( { 1 \over q^n } - 1 \right)  ,
\qquad
n = -1 , 0, 1, 2, ...
\end{multline}

Note that
$$
x_n \rightarrow
\left\{
\begin{array}{l}
\infty , \qquad 0 < q < 1  \\
\\
{\displaystyle { \tau   \over  q - 1 } }, \qquad q > 1 \\
\end{array}
\right.
\qquad
\mbox{as}
\qquad
n \rightarrow \infty   .
$$

In the particular case $[ x_{-1} , x_0 ] = [ - \tau , 0]$, we get
$$
x_n =  { \tau   \over 1 - q }   \left( { 1 \over q^n } - 1 \right)  ,
\qquad
n = -1 , 0, 1, 2, ...
$$

\item

For $ 0 <  q  < 1  $, $ \tau = 0 $ get
a particular sub{case} of case 2.
We get the $q$-delay relation
\begin{equation}   \label{qqmesh}
 x _-     =  q  x ,
\end{equation}
which is invariant with respect to the symmetry operator
$$
X_1  = x { \ddx} .
$$

For $ 0 < x_{-1} < x_0 $ this relation can be solved as
\begin{equation}   \label{q0mesh}
x_n =  {  x_0 \over q ^n } ,
\qquad
q=  {  x_{-1}  \over x_0  } ,
\qquad
n = -1, 0, 1, 2, ...
\end{equation}

\end{enumerate}

\subsection{Integration}

Let us show how the DODE~(\ref{most}) can be explicitly solved  
by the method of steps~\cite{Elgolts1964, bk:Myshkis1972} 
starting from the initial condition
$$
y(x) = \varphi (x) , \qquad x \in [ x_{-1} , x_0]  .
$$
We need to consider different cases of the delay relation~(\ref{mostmesh}).

\begin{enumerate}

\item   $q=1$.

The delay relation provides points~(\ref{uniform}).
The recursive relation
$$
y(x) = y(x_n )  e ^{ x - x_n \over  \tau }
-  e ^{ x / \tau } \int _{x_n} ^x    { 1 \over \tau }     e ^{ - { u  \over  \tau }  }      y (u - \tau) d u
$$
allows to find the solution on the interval $ [ x_{n} , x_{n+1} ] $
using the solution from the previous interval $ [ x_{n-1} , x_{n} ] $.

\item   $ q \neq 1$.

The delay relation gives points~(\ref{qmesh}).
We find the recursive relation
$$
y(x)
= y(x_n )  \left(   { (1 - q) x_n   + \tau  \over  (1 - q) x  + \tau } \right) ^{1 \over  q - 1}
$$
$$
-  \left(   { 1  \over  (1 - q)  x   + \tau } \right) ^{1 \over  q - 1}
 \int _{x_n} ^x    (   (1 - q) u  + \tau  ) ^{ 2 - q  \over  q - 1}   y ( q u  - \tau ) du  ,
$$
where $ \tau = q x_0 - x_{-1}$.

\item

$ 0 < q < 1 $ and  $\tau = 0 $ (sub{case} of case 2).

The delay relation gives points~(\ref{q0mesh}).
We find the recursive relation
$$
y(x)
= y(x_n )  \left(   {  x_n    \over   x  } \right) ^{1 \over  q - 1}
-    { 1  \over  (1 - q)  x   ^{1 \over  q - 1} }
 \int _{x_n} ^x     u    ^{ 2 - q  \over  q - 1}   y ( q u  ) du  ,
$$
where $  q = {\displaystyle  { x_{-1} \over x_0 } }$.

\end{enumerate}

\subsection{Symmetries with piecewise-smooth  coefficients}

The simplest linear  DODE~(\ref{most}) with a constant delay parameter
\begin{equation}     \label{primer}
 \dot{y} = { y - y _- \over x - x _- } ,
 \qquad
 x - x_- = const,
\end{equation}
is close to the second-order ODE $ \ddot{y}   = 0 $ in the sense of approximation order. 
Indeed, substituting the Taylor series
into~(\ref{primer}), we obtain
$$
{ \tau \over 2 } \left( \ddot{y}  - { \tau \over 3 } \dddot{y}  +
... \right) = 0.
$$
We suppose that $  x - x_- = \tau =  \mbox{const}  \ll  1 $.

Equation $ \ddot{y}  = 0 $ has general solution $ y = A x + B $,
which is a special solution of~(\ref{primer}).  
Meanwhile~(\ref{primer}) has infinitely many
{non}smooth solutions. In particular one can verify that it has the
following solution in a point $ x_0$
\begin{equation*}  \label{ini1}
 y (x) = (x- x_0 + \tau)^2 , \quad  x_0 -\tau \leq x \leq  x_0
\end{equation*}
$$
y(x) = -4\tau^2  e^{( x-x_0 ) /\tau}+ (x - x_0 + 2 \tau)^2 +\tau^2,
\quad x_0 \leq x \leq  x_0  + \tau
$$

Equation~(\ref{primer}) posses  in particular the following symmetry
\begin{equation}  \label{particularsym}
X = \alpha (x) { \partial \over   \partial y } , \qquad \dot{\alpha}
(x) = { \alpha (x)  -  \alpha (x_- )  \over x - x _- }.
\end{equation}
Substituting the solution given above into Lie equations, we can
present the following transformation group acting in the
neighborhood  of point $ x_0 $:
\begin{equation*}  \label{ini2}
y^*  (x) = y + (x- x_0 + \tau)^2 a  , \qquad x ^* = x ,\qquad  x_0
-\tau \leq x \leq  x_0,
\end{equation*}
$$
 y^*  (x) = y + \left(  -4\tau^2  e^{( x-x_0 ) /\tau}+ (x - x_0 + 2
\tau)^2 +\tau^2  \right) a ,\qquad x ^* = x , \qquad x_0 \leq x \leq
x_0 + \tau,
$$
where $a$  is a group parameter.

Thus, the transformation exists and it is defined by  non-smooth
function, which has a brake in a first derivative at  the point $x_0$.

It should be emphasized   
that these symmetries are not classical because
they are defined by piecewise smooth functions.

\section{Conclusions}

\label{Conclusions}

In this paper we considered first-order delay ordinary differential equations
which admit two linearly connected symmetries.
Consideration of a DODE requires an additional equation which specifies
the delayed argument $ x_- $. It was called the delay relation.
A DODE and a delay relation form a delay ordinary differential system. 
The obtained DODEs are all linear with a delay equation of the form $ x_- = g(x) $.

We provided a classification of all first-order DODSs specified by
two-, three- and four-dimensional Lie algebras
which contain two linearly connected symmetry operators.
These results are presented in Table~2.
It was noted that higher dimensional Lie algebras do not provide new DODSs.
However, the infinite-dimensional symmetry algebra of a DODS 
specified by a finite-dimensional {sub}algebra   of dimension $ 2 \leq n \leq 4 $ 
may contain finite-dimensional {sub}algebras of arbitrary higher dimension.

As an application of symmetry properties we have 
constructed exact analytic solutions of the obtained DODSs
with the help of symmetry reduction.
Such solutions, which are called invariant solutions,
were found using representative lists of one-dimensional {sub}algebras.

A point to emphasize is the role of the infinite-dimensional symmetry algebra 
$\mbox{\bf A}_{\infty}$ with basis~(\ref{always}). 
Similarly to the case of PDEs~\cite{Bluman1989}, 
the existence  of such an infinite-dimensional symmetry algebra for a nonlinear DODS indicates 
that this DODE is linearizable by a point transformation. This is an important result.

On the other hand the symmetry algebra (1.7) alone does not provide any invariant solutions, 
simply because it does not contain any elements of the form~(\ref{operator1}) with $\xi (x,y) {\not \equiv } 0$.  
In some cases the symmetry algebra  $\mbox{\bf A}_{\infty}$  
can be extended to a larger algebra $\tilde{\mbox{\bf A}}_{\infty}$  with $\mbox{\bf A}_{\infty}$ as an ideal. 
It was shown in~\cite{ourpaper}, Theorem~5.4 that at most one element with  $\xi  {\not \equiv } 0$ 
can be added to the basis of $\mbox{\bf A}_{\infty}$. 
Moreover that can only be done if the functions $\alpha (x)$, $\beta(x)$ and $g(x)$ 
in~(\ref{linearDODE}) satisfy a compatibility condition given in~\cite{ourpaper}. 
The additional element has the form
\begin{equation}
   Z = \xi (x)   {\partial \over \partial x }  + (A(x) y + B (x) )   {\partial \over \partial y }   ,  
\qquad    \xi (x)  {\not \equiv } 0   .                     
\end{equation}
Invariant solutions are obtained using elements of the form
\begin{equation}
    Z+pX,  \qquad   \mbox{ $X$   contained in  $\mbox{\bf A}_{\infty}$},  \quad    p = \mbox{const}.                   
\end{equation}
All such cases, up to equivalence are listed in Section~\ref{linearsec}.  
For instance,  we have $Z=   {\partial /  \partial x }$  
in $\mbox{\bf A}_{3,1}$, $\mbox{\bf A}_{3,13}$ and $\mbox{\bf A}_{4,12}$, 
 $ Z = (1-a) x  {\partial / \partial x } + y  {\partial / \partial y } $ for  $\mbox{\bf A}_{3,3}^a $,  
 $ Z={\partial / \partial x }  + y  {\partial / \partial y } $ for  $\mbox{\bf A}_{3,5}$, etc.

\bigskip

\noindent
{ \bf Acknowledgments}
The research of VD was partly supported by research grant No.~15-01-04677-A
of the Russian Fund for Base Research. 
The research of PW was partially supported by a research grant from NSERC of Canada.

\eject


\bigskip

\begin{center}
{\bf  Table~1. Lie symmetry algebras and their realizations containing two linearly connected vector fields.}
 \end{center}

In Column~1 we give the isomorphism class using the notations of~\cite{SWbook}.
Thus $ {\sso {n}}_{i,k} $ denotes the $k$-th nilpotent Lie algebra of dimension $i$ in the list.
The only nilpotent algebras in Table~1 are $ {\sso {n}}_{1,1} $, $ {\sso {n}}_{3,1} $ and $ {\sso {n}}_{4,1} $.
Similarly, $ {\sso {s}}_{i,k} $ is the $k$-th solvable Lie algebra of dimension $i$ in the list.
In Column~2 $ \mbox{\bf A}_{i,k} $ runs through all algebras in the list of sub{algebras}
of $ \mbox{diff} ( 2, \mathbb{R}  ) $ and $i$ is again the dimension of the algebra.
The simple Lie  algebra $ {\sso {sl}} ( 2, \mathbb{R}  ) $
is identified by its usual name.
The numbers in brackets correspond to notations used in the list of Ref.~\cite{Gonzalez1992}.
In Column~3 we give vector fields spanning each representative algebra.

Indecomposable Lie algebras precede the decomposable ones
(like  $ 2 {\sso {n}}_{1,1} $ or   $ {\sso {n}}_{1,1} \oplus  {\sso {s}}_{2,1} $)
in the list.
Isomorphic Lie algebras can be realized in more than one manner by vector fields.
For nilpotent Lie algebras elements of the derived algebra precede a semicolon,
e.g. $ X_1$,   $ X_2$ in    $ {\sso {n}}_{4,1} $.
For solvable Lie algebras the nil{radical} precedes a semicolon,
e.g. $ X_1$,   $ X_2$,    $ X_3$ in    $ {\sso {s}}_{4,6} $.

\bigskip
\bigskip

\noindent
{\bf Dimension 2}

$$
\begin{array}{|c|c|l|}
\hline
  &  &    \\
\mbox{Lie algebra} & \mbox{Case} & \mbox{Operators} \\
  &  &    \\
\hline
    &  &   \\
{\sso {s}} _{2,1}
&
\mbox{\bf A}_{2,1} (10)
&
{ \displaystyle
X_1 = { \ddy}   ;
\quad
X_2 =  y  { \ddy} }
\\
    &   &   \\
\hline
  &  &    \\
2 {\sso {n}} _{1,1}
&
\mbox{\bf A}_{2,3} (20)
&
{ \displaystyle
\left\{  X_1 = { \ddy}  \right\} ,
\quad
\left\{  X_2 = x  { \ddy}  \right\}  }
\\
    &      &   \\
\hline
\end{array}
$$

\eject

\noindent
{\bf Dimension 3}

$$
\begin{array}{|c|c|l|}
\hline
  &   &    \\
\mbox{Lie algebra} & \mbox{Case} & \mbox{Operators}  \\
  &   &    \\
\hline
  &   &    \\
{\sso {n}} _{3,1}
&
\mbox{\bf A}_{3,1} (22)
&
 { \displaystyle
X_1 = { \ddy} ;
\quad
X_2 =  x  { \ddy} ,
\quad
X_3 = { \ddx} }
\\
  &     &    \\
\hline
  &  &    \\
{\sso {s}} _{3,1}
&
\mbox{\bf A}_{3,3} ^a  (21, 22)
&
{ \displaystyle
X_1 = { \ddy} ,
\quad
X_2 = x { \ddy} ;
\quad
X_3 =  (1-a) x { \ddx}  +  y { \ddy} ,
\quad
0 < |a| \leq 1
  }
\\
  &    &   \\
\hline
   &  &   \\
{\sso {s}} _{3,2}
&
\mbox{\bf A}_{3,5}  (22)
&
 { \displaystyle
X_1 = { \ddy} ,
\quad
X_2 = x { \ddy} ;
\quad
X_3 =  { \ddx}  +  y { \ddy}   }
\\
   &     &   \\
\hline
   &  &   \\
{\sso {s}} _{3,3}
&
\mbox{\bf A}_{3,7}  ^b   (22)
&
{ \displaystyle
X_1 = { \ddy} ,
\quad
X_2 = x { \ddy} ;
\quad
X_3 =   ( 1   +  x ^2 )  { \ddx}   +  ( x + b)   y { \ddy}  ,
\quad
b \geq 0  }
\\
  &     &   \\
\hline
  &  &    \\
{\sso {sl}}  (2, \mathbb{R})
&
\mbox{\bf A}_{3,11}  (11)
&
 {\displaystyle
X_1 =   { \ddy} ,
\quad
X_2 = y  { \ddy} ,
\quad
X_3 = y^2  { \ddy}   }
\\
  &      &   \\
\hline
  &  &    \\
{\sso {n}}_{1,1}    \oplus {\sso {s}}_{2,1}
&
\mbox{\bf A}_{3,13}  (23)
&
{ \displaystyle
\left\{ X_1 = { \ddx} \right\} ,
\quad
\left\{
X_2 = { \ddy} ;
\quad
X_3 =  y { \ddy}
\right\}  }
\\
  &  &    \\
\cline{2-3}
 &  &    \\
&
\mbox{\bf A}_{3,14}   (22)
&
{ \displaystyle
\left\{  X_1 = x { \ddy}  \right\}   ,
\quad
\left\{
X_2 = { \ddy} ;
\quad
X_3 =   x { \ddx}  + y { \ddy}
\right\}   }
\\
  &   &    \\
\hline
  &  &    \\
3{\sso {n}}_{1,1}
&
\mbox{\bf A}_{3,15}  (20)
&
{\displaystyle
\left\{  X_1 =   { \ddy} \right\}  ,
\quad
\left\{  X_2 = x   { \ddy} \right\} ,
\quad
\left\{  X_3 =  \chi (x)   { \ddy}  \right\}  ,
\quad
\ddot{\chi}   (x) {\not\equiv}  0  }
\\
  &    &     \\
\hline
\end{array}
$$

\eject

\noindent
{\bf Dimension 4}


$$
\begin{array}{|c|c|l|}
\hline
  &   &     \\
\mbox{Lie algebra} & \mbox{Case} & \mbox{Operators}   \\
  &   &     \\
\hline
 &  &   \\
{\sso {n}}_{4,1}
&
\mbox{\bf A}_{4,1} (22)
&
  {\displaystyle
X_1 =   { \partial \over \partial y  } ,
\quad
X_2 =  x  { \partial \over \partial y  } ;
\quad
X_3 =  x ^2    { \partial \over \partial y  }   ,
\quad
X_4 = { \partial \over \partial x }
 }
\\
  &    &   \\
\hline
  &  &   \\
{\sso {s}}_{4,1}
&
\mbox{\bf A}_{4,2}  (22)
&
  {\displaystyle
X_1 =    { \partial \over \partial y  } ,
\quad
X_2 =  x    { \partial \over \partial y  } ,
\quad
X_3 =  e ^{x}   { \partial \over \partial y  } ;
\quad
X_4 = { \partial \over \partial x }
}
\\
  &    &   \\
\hline
 &  &   \\
{\sso {s}}_{4,2}
&
\mbox{\bf A}_{4,3}  (22)
&
  {\displaystyle
X_1 =   { \partial \over \partial y  } ,
\quad
X_2 =  x  { \partial \over \partial y  } ,
\quad
X_3 =  x ^2    { \partial \over \partial y  }  ;
\quad
X_4 = { \partial \over \partial x }  + y { \partial \over \partial y  }
 }
\\
  &   &   \\
\hline
 &  &   \\
{\sso {s}}_{4,3}
&
\mbox{\bf A}_{4,4}  ^{a, \alpha}   (22)
&
  {\displaystyle
X_1 =    { \partial \over \partial y  } ,
\quad
X_2 =  x { \partial \over \partial y  } ,
\quad
X_3 =   |x|  ^{ \alpha }   { \partial \over \partial y  }  ;
\quad
X_4 = ( 1 - a) x { \partial \over \partial x }  + y   { \partial \over \partial y } ,
}
\\
  &  &   \\
&  &
{\displaystyle
a \in [-1 , 0) \cup  (0,1)  , 
\ 
 \alpha  \neq  \{  0, {1 \over 1-a} , 1 \}   }
\\
  &  &   \\
&  &
\mbox{(see~\cite{SWbook} for additional restrictions on $a$  and $\alpha$)}
\\
  &   &   \\
\cline{2-3}
  &  &   \\
&
\mbox{\bf A}_{4,5} (21)
&
  {\displaystyle
X_1 = { \partial \over \partial y }   ,
\quad
X_2 =  x  { \partial \over \partial y  } ,
\quad
X_3 =  \chi (x) { \partial \over \partial y  } ;
\quad
X_4 = y { \partial \over \partial y }   ,
\quad
\ddot{\chi}(x) {\not\equiv} 0   }
\\
  &  &   \\
\hline
  &  &   \\
{\sso {s}}_{4,4}
&
\mbox{\bf A}_{4,6} ^a (22)
&
  {\displaystyle
X_1 =    { \partial \over \partial y  } ,
\quad
X_2 =  x    { \partial \over \partial y  } ,
\quad
X_3 =  e ^{ a x}   { \partial \over \partial y  } ;
\quad
X_4 = { \partial \over \partial x }  +  y { \partial \over \partial y  } ,
\quad
a \neq 0 , 1 }
\\
  &  &   \\
\hline
 &  &   \\
{\sso {s}}_{4,5}
&
\mbox{\bf A}_{4,7}  ^ {\alpha, \beta}  (22)
&
  {\displaystyle
X_1 =  { \partial \over \partial y  } ,
X_2 =  e ^{ \alpha  x} \cos ( \beta x )  { \partial \over \partial y  } ,
X_3 =  e ^{ \alpha x} \sin ( \beta x )  { \partial \over \partial y  } ;
X_4 = { \partial \over \partial x }   + y  { \partial \over \partial y  }  }, \\
  &  &   \\
 &  & 
 \beta \neq 0
  \\
  &   &   \\
\hline
\end{array}
$$

\eject

$$
\begin{array}{|c|c|l|}
\hline
  &   &     \\
\mbox{Lie algebra} & \mbox{Case} & \mbox{Operators}   \\
  &   &     \\
\hline
   &  &   \\
{\sso {s}}_{4,6}
&
\mbox{\bf A}_{4,8}  (24)
&
  {\displaystyle
X_1 =  { \partial \over \partial y }   ,
\quad
X_2 = { \partial \over \partial x }   ,
\quad
X_3 = x  { \partial \over \partial y  }  ;
\quad
X_4 =  x  { \partial \over \partial x  }
  }
\\
  &  &    \\
\hline
  &  &   \\
{\sso {s}}_{4,8}
&
\mbox{\bf A}_{4,9} ^a  (24)
&
  {\displaystyle
X_1 =  { \partial \over \partial y }   ,
\quad
X_2 = { \partial \over \partial x }   ,
\quad
X_3 = x  { \partial \over \partial y  }  ;
\quad
X_4 =  x  { \partial \over \partial x  }   +  a y  { \partial \over \partial y  } ,
\quad
a \neq  0,1
  }
\\
  &  &    \\
\hline

  &  &    \\
{\sso {s}}_{4,10}
&
\mbox{\bf A}_{4,10}  (25)
&
  {\displaystyle
X_1 =  { \partial \over \partial y }   ,
\quad
X_2 = { \partial \over \partial x }   ,
\quad
X_3 = x  { \partial \over \partial y  }  ;
\quad
X_4 =    x { \partial \over \partial x } +    ( 2 y + x^2 )  { \partial \over \partial y  } }
\\
  &   &    \\
\hline
   &  &   \\
{\sso {s}}_{4,11}
&
\mbox{\bf A}_{4,11} (24)
&
  {\displaystyle
X_1 =  { \partial \over \partial y }   ,
\quad
X_2 = { \partial \over \partial x }   ,
\quad
X_3 = x  { \partial \over \partial y  }  ;
\quad
X_4 =  x  { \partial \over \partial x  }   +  y  { \partial \over \partial y  }
  }
\\
  &  &    \\
\cline{2-3}
  &  &   \\
&
\mbox{\bf A}_{4,12} (23)
&
  {\displaystyle
X_1 =  { \partial \over \partial y }   ,
\quad
X_2 =  x  { \partial \over \partial y  } ,
\quad
X_3 =  { \partial \over \partial x  }  ; 
\quad
X_4 =  y { \partial \over \partial y  }
 }
\\
  &  &   \\
\hline
 &   &     \\
{\sso {s}}_{4,12}
&
\mbox{\bf A}_{4,14}   (23)
&
  {\displaystyle
X_1 =  { \partial \over \partial y  } ,
\quad
X_2 =  x  { \partial \over \partial y  } ;
\quad
X_3 =  y { \partial \over \partial y  } ,
\quad
X_4 = ( 1 + x^2 ) { \partial \over \partial x } +  x y { \partial \over \partial y }
  }
\\
  &   &   \\
\hline
\end{array}
$$

\eject

$$
\begin{array}{|c|c|l|}
\hline
  &   &     \\
\mbox{Lie algebra} & \mbox{Case} & \mbox{Operators}   \\
  &   &     \\
\hline
  &  &   \\
{\sso {n}}_{1,1} \oplus {\sso {s}}_{3,1}
&
\mbox{\bf A}_{4,15} ^a  (22)
&
  {\displaystyle
\left\{X_1 =   |x|  ^{1 \over 1-a}   { \partial \over \partial y  } \right\},
\left\{
X_2 =    { \partial \over \partial y  } ,
X_3 =  x { \partial \over \partial y  } ;
X_4 = ( 1 - a) x { \partial \over \partial x }  + y   { \partial \over \partial y }
\right\} ,
}
\\
  &  &   \\
&  &
{\displaystyle
a \in [-1 , 0) \cup  (0,1)   }
\\
  &   &   \\
\hline
  &  &   \\
{\sso {n}}_{1,1} \oplus {\sso {s}}_{3,2}
&
\mbox{\bf A}_{4,16} (22)
&
  {\displaystyle
\left\{ X_1 =  e ^{x}   { \partial \over \partial y  } \right\},
\quad
\left\{
X_2 =    { \partial \over \partial y  } ,
\quad
X_3 =  x    { \partial \over \partial y  } ;
\quad
X_4 = { \partial \over \partial x }  + y { \partial \over \partial y }
\right\}
  }
\\
  &  &   \\
\hline
  &  &   \\
{\sso {n}}_{1,1} \oplus {\sso {s}}_{3,3}
&
\mbox{\bf A}_{4,17} ^{ \alpha } (22)
&
  {\displaystyle
\left\{  X_1 = { \partial \over \partial y }   \right\}   ,
\left\{
X_2 =   { \partial \over \partial x  } ,
X_3 =  e ^{ \alpha  x} \cos  x   { \partial \over \partial y  } ,
X_4 =  e ^{ \alpha x}  \sin   x   { \partial \over \partial y  }
\right\} ,
} \\
  &   &   \\
  &   &  { \alpha  \geq 0 }
\\
  &   &   \\
\hline
 &  &   \\
{\sso {n}}_{1,1} \oplus {\sso {sl}}  (2, \mathbb{R})
&
\mbox{\bf A}_{4,18} (14)
&
  {\displaystyle
\left\{  X_1 = { \partial \over \partial x } \right\}   ,
\quad
\left\{
X_2 =  { \partial \over \partial y } ,
\quad
X_3 = y { \partial \over \partial y }  ,
\quad
X_4  = y^2 { \partial \over \partial y }
\right\}  }
\\
 &     &   \\
\cline{2-3}
 &   &   \\
&
\mbox{\bf A}_{4,19}  (19)
&
  {\displaystyle
\left\{ X_1 =   x { \partial \over \partial x } \right\} ,
\left\{
X_2 = { \partial \over \partial y }   ,
X_3 = x { \partial \over \partial x }  +   y { \partial \over \partial y } ,
X_4 = 2 x y   { \partial \over \partial x }  +  y^2   { \partial \over \partial y }
\right\}  }
\\
  &    &   \\
\hline
  &   &  \\
2 {\sso {s}}_{2,1}            
&
\mbox{\bf A}_{4,20} (13)
&
  {\displaystyle
\left\{
X_1 = { \partial \over \partial x } ;
\quad
X_2 = x { \partial \over \partial x }
\right\} ,
\quad
\left\{
X_3 =  { \partial \over \partial y }  ;
\quad
X_4  = y { \partial \over \partial y }
\right\} }
\\
  &   &   \\
\cline{2-3}
  &  &   \\
&
\mbox{\bf A}_{4,21}  (23)
&
  {\displaystyle
\left\{
X_1 =   { \partial \over \partial y  } ;
\quad
X_2 = x { \partial \over \partial x }  +   y { \partial \over \partial y  }
\right\} ,
\quad
\left\{
X_3 =  x { \partial \over \partial y  } ;
\quad
X_4 =  x { \partial \over \partial x  }
\right\} }
\\
  &   &   \\
\hline
 &  &   \\
4 {\sso {n}}_{1,1}
&
\mbox{\bf A}_{4,22}  (20)
&
  {\displaystyle
\left\{  X_1 = { \partial \over \partial y }  \right\}  ,
\left\{  X_2 = x { \partial \over \partial y  } \right\} ,
\left\{  X_3 =  \chi_1 (x) { \partial \over \partial y  } \right\} ,
\left\{  X_4 =  \chi_2 (x) { \partial \over \partial y  } \right\}  }
\\
  &  &   \\
&
&
\mbox{$1$, $x$, $\chi_1 (x)$ and $\chi_2 (x) $ are linearly independent}
\\
  &   &   \\
\hline
\end{array}
$$

\eject


\begin{center}
{\bf Table~2.  Invariant linear DODSs:
linear DODEs and delay relations, which do not depend on the solutions.}
\end{center}

\bigskip

\noindent
{\bf Dimension 2}

$$
\begin{array}{|c|c|c|}
\hline
    &   &  \\
\mbox{Case} & \mbox{DODE} & \mbox{Delay relation}  \\
    &   &  \\
\hline
     &   &  \\
\mbox{\bf A}_{2,1}
&
{ \displaystyle  \dot{y}   =    f (x)   {  \Delta y     \over    \Delta  x   } }
&
 x_-    =      g (x)
\\
    &   &  \\
\hline
  &   &  \\
\mbox{\bf A}_{2,3}
&
{ \displaystyle  \dot{y}  =   {  \Delta y    \over   \Delta  x  }    +  f (x)    }
&
   x_-    =     g ( x )
\\
    &   &  \\
\hline
\end{array}
$$

\eject

\noindent
{\bf Dimension 3}

$$
\begin{array}{|c|c|c|}
\hline
    &   &  \\
\mbox{Case} & \mbox{DODE} & \mbox{Delay relation}  \\
    &   &  \\
\hline
    &   &  \\
\mbox{\bf A}_{3,1}
&
 { \displaystyle   \dot{y}  =    { \Delta y  \over \Delta x  }   +  C_1 }
&
\Delta x  = C_2
\\
    &   &  \\
\hline
    &   &  \\
\mbox{\bf A}_{3,3} ^a  \quad a \neq 1
&
{ \displaystyle  \dot{y}  =    { \Delta y  \over \Delta x  }   +  C_1    x^{a \over 1-a}   }
&
 x_-     =  C_2  x
\\
    &   &  \\
\hline
    &   &  \\
\mbox{\bf A}_{3,3} ^a  \quad  a=1
&
 { \displaystyle    \dot{y}  =   { \Delta y  \over \Delta x  }    }
&
 x_-   = g (x)
\\
    &   &  \\
\hline
     &   &  \\
\mbox{\bf A}_{3,5}
&
 { \displaystyle   \dot{y}  =    { \Delta y  \over \Delta x  }   +  C_1   e^x  }
&
\Delta x  = C_2
\\
    &   &  \\
\hline
     &   &  \\
\mbox{\bf A}_{3,7} ^b
&
{ \displaystyle   \dot{y}  =    { \Delta y  \over \Delta x  }
+  C_1   { e^{ b   \arctan (x) }  \over  \sqrt{ 1 + x^2 }  }   }
&
{ \displaystyle  x_- = { x   - C_2   \over   1 +   C_2  x }    }
\\
    &   &  \\
\hline
    &   &  \\
\mbox{\bf A}_{3,11}
&
\mbox{No DODE}
&
\\
    &   &  \\
\hline
    &   &  \\
\mbox{\bf A}_{3,13}
&
{ \displaystyle    \dot{y}  =    C_1 { \Delta y  \over \Delta x  }    }
&
\Delta x  =  C_2
\\
    &   &  \\
\hline
    &   &  \\
\mbox{\bf A}_{3,14}
&
{ \displaystyle   \dot{y}  =    { \Delta y  \over \Delta x  }   +  C_1   }
&
x_-     =  C_2  x
\\
    &   &  \\
\hline
    &   &  \\
\mbox{\bf A}_{3,15}
&
{\displaystyle   \dot{y} =  { \Delta y \over \Delta x  } + f(x)  }
&
{\displaystyle  \dot{\chi}(x)   = { \chi (x) - \chi ( x_-)  \over  x  - x_-   }  }
\\
    &   &  \\
\hline
\end{array}
$$

\eject

\noindent
{\bf Dimension 4}

$$
\begin{array}{|c|c|c|}
\hline
    &   &  \\
\mbox{Case} & \mbox{DODE} & \mbox{Delay relation}  \\
    &   &  \\
\hline
   &   &  \\
\mbox{\bf A}_{4,5}
&
{\displaystyle   \dot{y} =  { \Delta y \over \Delta x  } }
&
{\displaystyle  \dot{\chi} (x)   = { \chi (x) - \chi ( x_-)  \over  x  - x_-   }  }
\\
  &   &  \\
\hline
    &   &  \\
\mbox{\bf A}_{4,12}
&
 {\displaystyle   \dot{y}   =  {  \Delta   y   \over \Delta x }    }
&
   \Delta  x     =   C
\\
  &   &  \\
\hline
  &   &  \\
\mbox{\bf A}_{4,14}
&
  {\displaystyle   \dot{y}   =  {  \Delta   y   \over \Delta x }    }
&
 {\displaystyle  x_- = { x   - C   \over   1 +    C x }    }
\\
  &   &  \\
\hline
  &   &  \\
\mbox{\bf A}_{4,21}
&
 {\displaystyle   \dot{y}   =  {  \Delta   y   \over \Delta x }     }
&
     x _-     = C   x
\\
  &   &  \\
\hline
\end{array}
$$


\begin{thebibliography}{10}




\bibitem{ourpaper}
V.~A.~Dorodnitsyn,  R.~Kozlov, S.~V.~Meleshko and  P.~Winternitz,
Lie group classification of first-order delay ordinary differential equations, 
arXiv:1712.02581






\bibitem{Ovsiannikov1982}
L.~V.~Ovsiannikov, 
{\it Group Analysis of Differential Equations},   
(Academic, New York, 1982)





\bibitem{Olver1986} 
P.~J.~Olver, 
{\it Applications of Lie Groups to Differential Equations},  
(Springer, New York, 1986)




\bibitem{Ibragimov1985}
N.~H.~Ibragimov, 
{\it Transformation Groups Applied to Mathematical Physics},  
(Reidel, Boston, 1985)





\bibitem{Bluman1989}
G.~W.~Bluman and S.~Kumei, 
{\it Symmetries and Differential Equations}, 
(Springer, Berlin, 1989)





\bibitem{Gaeta1994} G. Gaeta, 
{\it Nonlinear Symmetries and Nonlinear Equations}, 
(Kluwer, Dordrecht, 1994)





\bibitem{DKW2000}
V.~Dorodnitsyn, R.~Kozlov and P.~Winternitz,  
 Lie group classification of second-order ordinary difference equations,
{\it J. Math. Phys. } {\bf  41}(1) 480--504  (2000)



\bibitem{Dorodnitsyn2011}
V.~Dorodnitsyn, 
{\it Applications of Lie Groups to Difference Equations}, 
(Chapman \& Hall/CRC differential and integral equations series, 2011)



\bibitem{Levi2006}
D.~Levi and P.~Winternitz 
(2006) 
Continuous symmetries of difference equations
{\it  J. Phys A } {\bf 39}  R1--R63





\bibitem{Elgolts1964} L.~E.~El'sgol'c,   
{\it Qualitative Methods in Mathematical Analysis},  
(GITTL, Moscow, 1955). 
American Mathematical Society (Translations
of Monographs Reprint), 1964.


\bibitem{bk:Myshkis1972}
A.~D.~Myshkis,  
{\it Linear differential equations with retarded argument},  
(Nauka, Moscow, 1972)







\bibitem{Gonzalez1992}
A.~Gonzalez-Lopez, N.~Kamran, and P. ~J.~Olver,
Lie algebras of vector fields in the real plane, 
{\it Proc. London Math. Soc. } {\bf 64}, 399  (1992)





\bibitem{SWbook}
L.~$\check{\mbox{S}}$nobl and P.~Winternitz,  
{ \it Classification and identification of Lie algebras},
CRM Monograph Series  33, American Mathematical Society,  (Providence, RI, 2014)





\bibitem{PateraWinternitz}
J.~Patera and P.~Winternitz,
{Sub}algebras of real three- and four-dimensional Lia algebras,
{\it  J.~Math.~Phys.}  {\bf  18},    1449--1455  (1977)





\bibitem{Gelfond}  A.~O.~Gelfond,  
{\it Calculus of finite differences}, 
International Monographs on Advanced Mathematics and Physics, 
(Delhi,  Hindustan Publishing Corp., 1971)



\bibitem{Jordan}  Ch.~Jordan,   
{\it Calculus of finite differences}, 
Third Edition, 
(New York,  Chelsea Publishing Co., 1965)
























\end{thebibliography}
\end{document}